\documentclass[submit,links=false,lastpage=false]{aiaa-two}
\usepackage{graphicx}
\usepackage{amsmath}
\usepackage{amsthm}
\usepackage{amsfonts}
\usepackage[labelformat=simple]{subfig}
\usepackage{epsf}
\usepackage{epsfig}

\author[Armanious and Lind]{ %
George Armanious\thanks{Graduate Student, Department of
	Mechanical and Aerospace Engineering, {\em a1george@ufl.edu}}
	\:\: and \:\:
Rick Lind\thanks{Associate Professor, Department of Mechanical
	and Aerospace Engineering, {\em ricklind@ufl.edu}}\\
\textit{University of Florida, Gainesville, FL 32612}}

\title{{Efficient Numerical Analysis of Stability} \\
	{of High-Order Systems With a Time Delay}}

 \abstract{ %
  Time delays are a common perturbation in systems with many states, such as networked, distributed, or decentralized systems.  Current methods analyzing the stability of large systems with time delay typically produce very conservative results.  While more exact methods exist, these become inefficient for large systems.  This paper provides a methodology for analyzing the stability of time-delayed systems that is derived from exact methods but is efficient for high-order systems.  The computational and memory cost of this new technique is compared to the costs of existing techniques, and its efficiency is shown using a distributed system with over four hundred states}

\begin{document}
\maketitle

\section*{Nomenclature}
\noindent\begin{tabular}{@{}lcl@{}}
	$A_0$ &=& Undelayed state matrix \\
	$A_1$ &=& Delayed state matrix \\
	$x$ &=& System state vector \\
	$t$ &=& Time \\
	$\tau$ &=& Time delay \\
	$\omega$ &=& Natural frequency at root crossing \\
	$T$ &=& Rekasius substitution variable \\
	$n$ &=& Number of system states \\
	$s$ &=& Frequency domain variable \\
	$I_n$ &=& $n\times n$ identity matrix \\
\end{tabular} \\

\section{Introduction}

A time delay of $\tau \in \mathcal{R}^+$ notes that dynamics of a system
at time of $t \in \mathcal{R}^+$ depend
on some state or input at a previous time of $t-\tau$.  Such a delay
can often arise due to hardware associated with sensing, actuation,
and communication.  It can be challenging to properly account for
a time delay when designing a control system; consequently, the
time delay can introduce unexpected instabilities.

This paper considers a special case of time-delayed systems known
as retarded time-delayed systems~\cite{139}.  This case of system,
shown in Eq.~\ref{eq:dynamics}, relates a state vector of $x\in\mathcal{R}^n$
to state matrices of $A_0,A_1 \in \mathcal{R}^{n \times n}$.  The
classification of retarded notes the absence of any $\dot{x}(t-\tau)$ term.

\begin{equation}
\dot{x}(t)=A_0x(t)+A_1x\left(t-\tau \right)
\label{eq:dynamics}
\end{equation}

The values of $\tau$ at which the system changes between stable
and unstable relate to the poles of the dynmics, or the roots
of the determinant in Eq.~\ref{eq:stability}, being purely imaginary.
A finite number of purely-imaginary roots exist but each root
actually corresponds to an infinite number of time delays because of
the transcendental term.


\begin{equation}
\det\left[sI_n-A_0-A_1e^{\tau s}\right] = 0
\label{eq:stability}
\end{equation}

All the time delays at which the system changes stability are actually
computed using a single $\tau$ and pole value~\cite{131}.  Essentially,
the condition for any pole crossing the imaginary axis depends on
the frequency of that pole at that crossing instead of the time
delay associated with that crossing.  This relationship between
all the time delays associated with changing stability is given
in Eq.~\ref{eq:tau0} for a pole of $s=\jmath \omega_i$ and the
set of time delays of $\tau_0,\tau_1,\tau_2,\hdots$  associated with
changes in stability.  Note that the system may have several poles
that cross the imaginary axis at different frequencies so the
stability analysis may want to compute all the sets of $\omega_i$
and associated $\tau$ values.

\begin{equation}
\tau_k = \tau_0+\frac{2\pi}{\omega_i}k,\>\>\>\>k=0,1,2,...
\label{eq:tau0}
\end{equation}



A symbolic method is formulated to find all sets of frequencies
and time delays associated with change in stability~\cite{131,133}.  This
method generates an algebraic expression for the characteristic
polynomial in terms of $s$ and $e^{-\tau s}$ and then utilizes the
the Rekasius substitution to replace the transcendental exponential
with a fraction of rational polynomials~\cite{rekasius}.  The Routh's
array is then used to find the conditions at which stability changes.  The
method generates an accurate set of conditions; however, the array
is exceptionally difficult to formulate and solve as the number
of states in the system increases beyond a few~\cite{142}.

A numerical method is also formulated by introducing Kronecker multiplication
to recast the $n \times n$ transcendental determinant as a $2n^2 \times
2n^2$ eigenvalue problem~\cite{139,140}.  These eigenvalues indicate all the poles
for which the stability changes and are used to find the associated
time delays.  The computations use standard tools for eigenvalues;
however, the memory storage and processing time make the approach
nearly infeasible for systems with more than 100 states.

Other approaches are formulated for decentralized systems~\cite{135,136}.
These approaches construct a Lyapunov function that generates an
upper bound on the time delay that can be tolerated before the onset
of instability.  These approaches are straightforward to apply
to any system and the computations are not overly burdensome; however,
some function parameters are challenging to choose and the
upper bound is often overly conservative.

This paper introduces an alternative method that combines elements
of the Rekasius-based approach and the Kronecker-based approach.  In
this alternative method, the Rekasius substitution is again used
but it is introduced earlier in the formulation than in the
previous method.  The new formulation is then converted 
from a determinant computation to an eigenvalue computation
involving a companion matrix in a similar process as used by
the previous method.  The resulting formulation is solved
numerically using a reasonable amount of memory and computations.

\section{Existing Methods}\label{existing}

\subsection{Rekasius Substitution} \label{sub:rek}
The characteristic equation of the system in Equation~\ref{eq:dynamics} is shown in Equation~\ref{CEStau}.  
\begin{equation}
\det\left[sI_n-A_0-A_1e^{\tau s}\right] = 0
\label{CEStau}
\end{equation}
The transcendental term makes the system infinite dimensional with infinitely many characteristic roots.  The determinant can be carried out symbolically to yield the general form shown in Equation~\ref{CESex}, where $a_k(s)$ are polynomials of degree $n-k$.
\begin{equation}
\det\left[sI_n-A_0-A_1e^{\tau s}\right] = \sum_{k=0}^{n} a_k(s)e^{-k\tau s}
\label{CESex}
\end{equation}
The foundation for the algorithm is the Rekasius substitution shown
in Eq.~\ref{eq:rekasius} for a delay of $\tau \in \mathcal{R}^+$ and
an associated $T \in \mathcal{R}$ and $s \in \mathcal{C}$.  The
substitution allows the transcendental term of a complex exponential
to be replaced with a fraction of complex polynomials~\cite{rekasius}.  The
critical issue for this substitution is its validity is restricted to
purely-imaginary values of $s$ such that $s=\jmath \omega$ for
$\omega \in \mathcal{R}$; consequently, considering $s$ as a pole
of the system implies the substitution is valid for crossings of
a pole between the left-half plane and the right-half plane which
are associated with neutral stability.

\begin{equation}
e^{-\tau s} = \frac{1-Ts}{1+Ts}
\label{eq:rekasius}
\end{equation}
The mapping from $T$ to $\tau$ is one-to-many for a given $\omega$ and is shown in Equation~\ref{substitution}.
\begin{equation}
\tau = \frac{2}{\omega}\left[\tan^{-1}(\omega T) \mp \ell \pi \right],\>\>\>\>\ell=0,1,2,...
\label{substitution}
\end{equation}
Applying the Rekasius substitution into Equation~\ref{CESex} yields Equation~\ref{CESfrac}.
\begin{equation}
\sum_{k=0}^{n} a_k(s)\left(\frac{1-Ts}{1+Ts}\right)^k = 0
\label{CESfrac}
\end{equation}
By multiplying both sides of Equation~\ref{CESfrac} by $(1+Ts)^n$, the equation can be simplified to  Equation~\ref{CETsPoly1}.  
\begin{equation}
\sum_{k=0}^n a_k(s)(1+Ts)^{n-k}(1-Ts)^k=0
\label{CETsPoly1}
\end{equation}
Equation~\ref{CETsPoly2} is obtained by grouping the terms in Equation~\ref{CETsPoly1} by powers of $s$, where $b_k(T)$ are polynomials of $T$.
\begin{equation}
\sum_{k=0}^{2n}b_k(T)s^k =0
\label{CETsPoly2}
\end{equation}

The polynomial in Equation~\ref{CETsPoly2} has the same roots as the polynomial of Equation~\ref{CESex} but does not contain an exponential in $s$.  Rather, each coefficient of $s$ is parameterized by a polynomial in $T$.  A Routh's array can be constructed for the polynomial of Equation~\ref{CETsPoly2}, and the first column of the array can be used to determine the pairs, $(T,s)=(T_c,\jmath\omega_c)$, that are roots of Equation~\ref{CETsPoly2}, and Equation~\ref{CESex} by extension.  

Finally, the time delay at which the dynamics in Eq.~\ref{eq:dynamics}
become unstable are computed for those values of $T$ at which
some eigenvalues are purely real.  This time delay of $\tau$ results
from the values of $s=\jmath \omega$ and $T$ that satisfy
Eq.~\ref{CETsPoly2} using the equivalence in Eq.~\ref{eq:tau}.

\begin{equation}
\tau = \frac{2}{\omega}\left[\tan^{-1}(\omega T) \mp \ell \pi \right],\>\>\>\>\ell=0,1,2,...
\label{eq:tau}
\end{equation}

Each value of $T$ and corresponding value of $\omega$ are associated with an infinite number of time delays periodically spaced by $\frac{2\pi}{\omega}$.  The transcendality in the original dynamics implies there are infinitely many roots in the characteristic equation, but literature shows these roots are associated with only a finite number of frequencies.  The Rekasius substitution introduces a new variable, $T$, and shifts the transcendality from the dynamics of the system to the relation between $T$ and the associated time delays, $\tau$, at frequencies, $\omega$, where the substitution is valid.  Literature also shows that the direction of a pole crossing at a particular time delay is only a property of the associated frequency.  In other words, all the values of $\tau$ related to a particular pair of $T$ and $\omega$ by Eq.~\ref{eq:tau} are either stabilizing or destabilizing~\cite{131}.

Two properties of a Routh's array can be taken advantage of to reduce the necessary computations in determining the purely imaginary roots, $s=\jmath\omega_c$, and corresponding parameter values, $T_c$, of Equation~\ref{CETsPoly2} \cite{133}.  The first property is that if there is a pair of imaginary roots, the only term on the row corresponding to $s^1$, defined here as $R_1(T)$, must be zero.  This property converts analyzing the entire first column into a single root finding problem in $T$.  Also, only purely real values of $T$ are of interest, since the Rekasius substitution is only defined for $T\in \mathcal{R}$.  Define $T_c$ as a real root of $R_1(T)$.

The second helpful property involves the row of corresponding to $s^2$.  This row has exactly two elements.  Define $R_{21} (T)$ and $R_{22}(T)$ as the first and second elements of this row.  Then for $T_c$, $\left(R_{21}(T_c)s^2+R_{22}(T_c)\right)$ is a factor of the characteristic equation in Equation~\ref{CETsPoly2}.  Therefore a subset of the roots of the original characteristic equation can be found by solving Equation~\ref{auxForCE} for $s$.  

\begin{equation}
R_{21}(T_c)s^2+R_{22}(T_c) = 0
\label{auxForCE}
\end{equation}

In order for the roots of Equation~\ref{auxForCE} to be purely imaginary, $s=\jmath\omega_c$, the product of $R_{21}(T_c)$ and $R_{22}(T_c)$ must be positive.  If this condition is met, the corresponding root crossing is given by Equation~\ref{omegaCSolved}.

\begin{equation}
\omega_c = \sqrt{\frac{R_{22}(T_c)}{R_{21}(T_c)}}
\label{omegaCSolved}
\end{equation}

An alternative method involving sum of squares optimization can be used to avoid constructing a Routh's array from Equation~\ref{CETsPoly2} in order to find the upper bound of the first stable time delay region \cite{142}.  This method searches through all values of $T$ and $s=\jmath\omega$ to find the largest value of $T$ and smallest value of $\omega$ such that the magnitude of Equation~\ref{CETsPoly2} is minimized.  A sum of squares algorithm is used to efficiently find these values of $T$ and $\omega$ given the characteristic equation.

\subsubsection{Advantages}
The main advantage of this method is its exactness.  The root crossings obtained from the Routh's array are the locations where the original system will cross the imaginary axis.  This method also provides the direction that the poles move across the imaginary axis at these root crossings and the time delays associated with each root.  From all of this information, a complete stability analysis can be performed which may reveal multiple regions of stability.  Of course, if one is only interested in the smallest time delay that causes instability, only the smallest time delay produced by this method is needed, as this time delay is guaranteed to make the system unstable, assuming the delay-free system is stable.  

\subsubsection{Disadvantages}
The key disadvantage is that this method is symbolic.  In the beginning of the formulation, the computation of a characteristic equation in terms of two symbolic variables, $s$ and $e^{-\tau s}$, is required.  Then, the Rekasius substitution is used to convert this symbolic equation into an equation in terms of $s$ and a new variable, $T$.  Finally, a Routh's array is constructed from this symbolic characteristic equation.  While only the final terms of the array are necessary to find the imaginary roots, the whole array must be constructed to obtain these terms.  For a small system, the computational efficiency of this method is comparable to a numerical method, but for a large system (i.e. a closed-loop distributed system with multiple controllers and observers), computing the characteristic equation can be tedious or infeasible depending on the strategy used.  

The sum of squares approach does not require construction of a Routh's array.  However, it assumes that the characteristic equation is symbolically computed for use in the sum of squares algorithm.  As mentioned, computing the characteristic equation of a large system analytically is infeasible.

\subsection{Kronecker Multiplication} \label{sub:kron}
The next method is a numerical approach to finding the imaginary root crossings that recasts the $n\times n$ transcendental determinant as a $2n^2 \times 2n^2$ eigenvalue problem in terms of the system matrices that can be solved numerically \cite{139,140}.  To begin, it is necessary to define two matrix operations.  First, the operation $\xi: C^{n\times n} \rightarrow C^{n^2\times 1}$ converts the square $ n\times n$ matrix, $M=\left[m_1^T, m_2^T, ..., m_n^T\right]^T$ into a $n^2 \times 1$ column matrix as shown in Equation~\ref{xiMMM}.
\begin{equation}
\left(\xi M\right)_{n^2\times 1} := \begin{bmatrix} m_1^T\\m_2^T\\\vdots\\m_n^T \end{bmatrix}
\label{xiMMM}
\end{equation}
The Kronecker product is also used in this formulation and is defined for any two matrices, $A_{n\times m}$ and $B_{p\times q}$ as shown in Equation~\ref{kronDef}.
\begin{equation}
A \otimes B = \begin{bmatrix} a_{11}B & \cdots & a_{1m}B \\ \vdots & \ddots & \vdots \\ a_{n1}B & \cdots & a_{nm}B \end{bmatrix}_{np \times mq}
\label{kronDef}
\end{equation}
The product of three $n \times n$ matrices, $P_1$, $P_2$, and $P_3$ can be turned into a Kronecker product of dimension $n^2 \times n^2$ using the $\xi$ operation and property shown in Equation~\ref{fancyPants} \cite{BernsteinMatrix}.
\begin{equation}
\xi(P_1P_2P_3) = (P_3^T\otimes P_1)\xi P_2
\label{fancyPants}
\end{equation}
This method begins with the original time-delayed ODE, shown again in Equation~\ref{retarded}.  
\begin{equation}
\dot{x}=A_0x+A_1x\left(t-\tau \right)
\label{retarded}
\end{equation}
The solution is assumed to have the form shown in Equation~\ref{ass}, where $v$ is a vector and $s$ is a scalar.  
\begin{equation}
x(t) = e^{st}v
\label{ass}
\end{equation}
The derivative of the assumed solution is shown in Equation~\ref{assDot}.
\begin{equation}
\dot{x}(t) = se^{st}v
\label{assDot}
\end{equation}
Substituting Equations~\ref{ass} and~\ref{assDot} into Equation~\ref{retarded} yields Equation~\ref{retardedAssPlug}.
\begin{equation}
se^{st}v = A_0e^{st}v+A_1e^{s(t-\tau)}v
\label{retardedAssPlug}
\end{equation}
Rearranging this equation and factoring $e^{st}$ and $v$ yields Equation~\ref{retardedAssPlug2}.
\begin{equation}
e^{st}(sI_n - A_0-A_1e^{-s\tau})v = 0
\label{retardedAssPlug2}
\end{equation}
If $s=\omega \jmath$ is a root, so is $-s=-\omega\jmath$.  The transpose of Equation~\ref{retardedAssPlug2} is taken to get Equation~\ref{retardedAssPlugT}.
\begin{equation}
v^*(-sI_n-A_0^T)=e^{\tau s}v^* A_1^T
\label{retardedAssPlugT}
\end{equation}
Multiply Equations~\ref{retardedAssPlug2} and~\ref{retardedAssPlugT} to get Equation~\ref{retardedAssPlugM}.
\begin{equation}
\left(sI_n-A_0\right)v\left(-v^*\right)\left(sI_n+A_0^T\right) = \left(e^{-\tau s}A_1v\right)\left(e^{\tau s} v^*A_1^T\right)
\label{retardedAssPlugM}
\end{equation}
Define $V$ as the outer product of $v$ with itself: $V:=vv^*$.  Substituting $V$ into Equation~\ref{retardedAssPlugM} yields Equation~\ref{retardedAssPlugV}.
\begin{equation}
\left(sI_n-A_0\right)V\left(sI_n+A_0^T\right) = -A_1VA_1^T
\label{retardedAssPlugV}
\end{equation}
Taking the transpose of both sides of Equation~\ref{retardedAssPlugV} yields Equation~\ref{retardedAssPlugTranspose}.
\begin{equation}
\left(sI_n+A_0\right)V\left(sI_n-A_0^T\right) = -A_1VA_1^T
\label{retardedAssPlugTranspose}
\end{equation}
Using the $\xi$ operator on both sides of the equation and applying the property in Equation~\ref{fancyPants} yields Equation~\ref{ximera}.
\begin{equation}
\left[\left(sI_n-A_0^T\right)^T\otimes\left(sI_n+A_0\right)\right]\xi V = -\left[\left(A_1^T\right)^T\otimes A_1\right]\xi V
\label{ximera}
\end{equation}
Rearranging Equation~\ref{ximera} yields Equation~\ref{chimera}.
\begin{equation}
\left[\left(sI_n-A_0\right)\otimes\left(sI_n+A\right)+A_1\otimes A_1\right] \xi V = 0
\label{chimera}
\end{equation}
Define $\lambda(s)$ as the coefficient of $\xi V$ in Equation~\ref{chimera} and rearrange the terms in powers of $s$ as shown in Equation~\ref{lambdaGROUP}.
\begin{equation}
\lambda(s)=s^2(I_n\otimes I_n) + s((I_n \otimes A_0) - (A_0\otimes I_n)) + ((A_1\otimes A_1)-(A_0\otimes A_0))
\label{lambdaGROUP}
\end{equation}
Equation~\ref{chimera} simplifies to $\lambda(s)\xi V = 0$.  The nontrivial solution of this equation is obtained by setting $\det (\lambda(s)) = 0$.  Note that $\lambda(s)$ is a matrix polynomial, as shown in Equation~\ref{poly}.
\begin{equation}
\lambda \left(s\right) = G_0s^2 + G_1s + G_2
\label{poly}
\end{equation}
The coefficients $G_0$, $G_1$, and $G_2$ are defined in Equation~\ref{g0g1g2}.
\begin{equation}
\begin{array}{c}
G_0 := \left(I_n \otimes I_n\right) = I_{n^2}\\
G_1 := ((I_n \otimes A_0) - (A_0 \otimes I_n)) \\
G_2 := ((A_1 \otimes A_1) - (A_0 \otimes A_0)) \\
\end{array}
\label{g0g1g2}
\end{equation}
The goal of setting the determinant of $\lambda(s)$ to zero can be achieved by linearlizing the polynomial \cite{matrixPoly}.  The companion matrix of Equation~\ref{poly} is defined in Equation~\ref{companion}.  The eigenvalues of this companion matrix are the values of $s$ that make the determinant of Equation~\ref{poly} equal to zero.
\begin{equation}
C := \begin{bmatrix} 0_{n^2} & I_{n^2}\\ -G_0^{-1}G_2 & -G_0^{-1}G_1 \end{bmatrix} = \begin{bmatrix} 0_{n^2} & I_{n^2}\\ -G_2 & -G_1 \end{bmatrix}
\label{companion}
\end{equation}
It can be easily shown that the determinant of $\lambda(s)$ is equal to that of $sI-C$, as shown in Equations~\ref{companionProof}-\ref{companionProofEnd}.  
\begin{eqnarray}
\det(sI_{2n^2}-C)& = & \det\left(sI_{2n^2}-\begin{bmatrix} 0_{n^2} & I_{n^2} \\
 -G_2 & -G_1 \end{bmatrix} \right) \label{companionProof}\\
& = &\det\left(\begin{bmatrix} sI_{n^2} & -I_{n^2} \\ G_2 & sI_{n^2}+G_1 \end{bmatrix}\right) \\
& = &\det\left(I_{n^2}s^2+G_1s+G_2\right) \\& = & \det(\lambda(s)) \label{companionProofEnd}
\end{eqnarray}

With this property, the numerically intensive problem of finding the imaginary root crossings of the system in Equation~\ref{retarded} simplifies to determining the eigenvalues of $C$, for which there are many existing algorithms.  Once all of the imaginary roots are found, the corresponding initial time delay can be determined.  For an imaginary root, $s=\jmath\omega$, the stability matrix, $T=\jmath\omega I_n - A_0 - A_1e^{-\jmath\omega\tau}$ is only singular if $|e^{-\jmath\omega\tau}|=0$.  The values of $z:=e^{-\jmath\omega\tau}$ that cause $T$ to be singular are the generalized eigenvalues of the pair $(A_0-\jmath\omega I_n, -A_1)$.  Let $z_d$ be the generalized eigenvalue that satisfies $|z_d|=1$.  Then the initial time delay corresponding to an imaginary axis crossing at $s=\jmath\omega$ is the smallest positive delay, $\tau$, that satisfies Equation~\ref{tauInitEig}, where $k$ is an integer and imag$(z_d)$ and real$(z_d)$ represent the imaginary and real components of $z_d$, respectively.
\begin{equation}
\tau = \frac{1}{\omega}\left(\tan^{-1}\left(\frac{-\mathrm{imag}(z_d)}{\mathrm{real}(z_d)}\right) + 2\pi k\right)
\label{tauInitEig}
\end{equation}
\subsubsection{Advantages}
This approach is very easy to formulate in code.  Given the matrices, $A_0$ and $A_1$, one only need compute the matrix $C$ given by Equation~\ref{companion} using the block matrices obtained by Equation~\ref{g0g1g2}.  In MATLAB, the kron command can be used to compute the Kronecker products used by Equation~\ref{g0g1g2}.  The eigenvalues of $C$ can be computed to obtain all of the imaginary axis crossings, and for each crossing, a smaller generalized eigenvalue problem is used to compute the corresponding time delay.  Both eigenvalue computations can be performed using the eig command in MATLAB.
\subsubsection{Disadvantages}
\begin{figure}[htbp]
	\centering
	\includegraphics[width=4in]{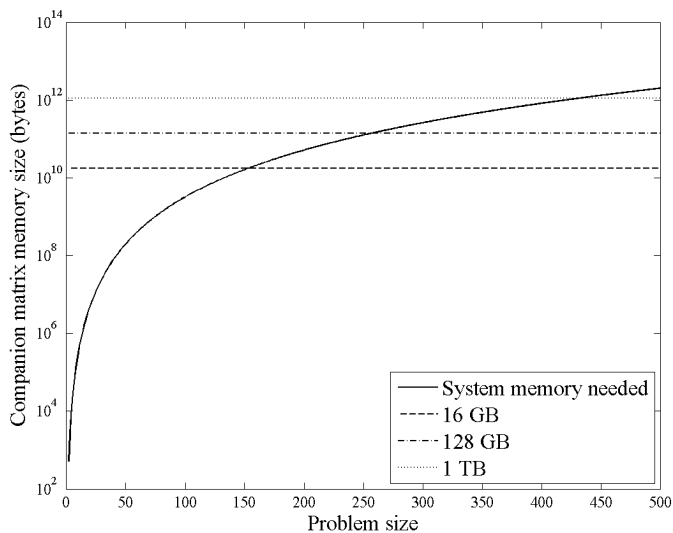}
	\caption[Memory Usage by Kronecker Multiplication Method]{System memory needed to store companion matrix in Kronecker multiplication method}
	\label{memNeedEig}
\end{figure}

For large state-space problems ($n\geq 150$), the companion matrix, $C$, from Equation~\ref{companion} becomes very large.  Each block element of $C$ has dimensions $n^2 \times n^2$ as a result of the Kronecker product.  Therefore, the total dimensions of $C$ are $2n^2 \times 2n^2$.  Storing such a matrix in memory becomes an issue for large $n$, and the eigenvalue operation, which has a theoretical time complexity of $\sim O(n^3)$, takes exponentially longer to compute.  For an example of how much system memory is required, consider a state-space size of 200.  The corresponding companion matrix has dimensions $80,000 \times 80,000$, and therefore contains 6.4~billion elements.  If each element requires 8 bytes of memory to store (the standard size in MATLAB), 51.2~GB of system memory is required to store this companion matrix!  The majority of modern consumer grade computers have much less RAM (typically between 8~GB and 16~GB), and recall that the starting state-space size is 200 states.  Figure~\ref{memNeedEig} shows required system memory to store the companion matrix as a function of starting state-space size.  

\begin{figure}[htbp]
	\centering
	\includegraphics[width=4in]{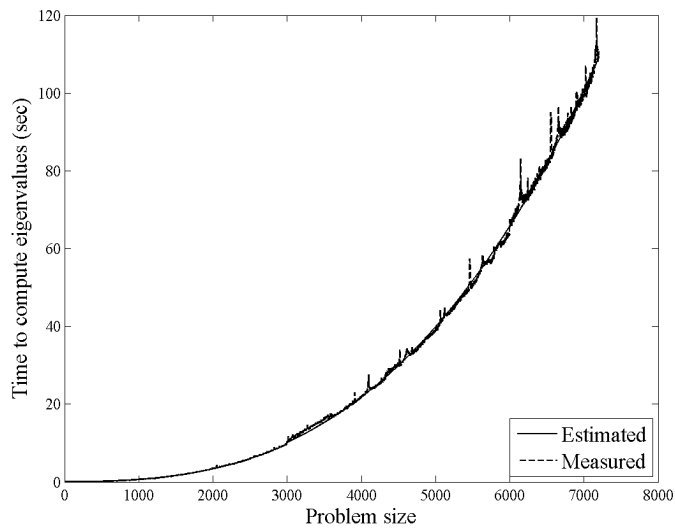}
	\caption[CPU time for eigenvalue calculations]{Measured and curve-fitted CPU time necessary for eigenvalue calculations}
	\label{cpuEig}
\end{figure}

\begin{figure}[htbp]
	\centering
	\includegraphics[width=4in]{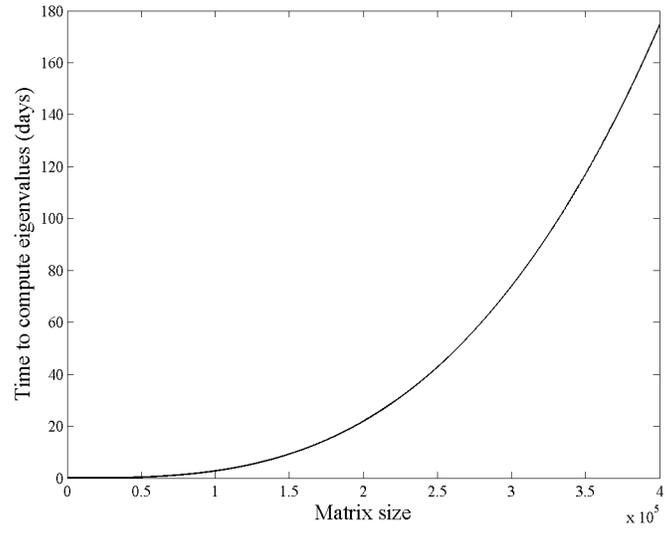}
	\caption[Estimated CPU time for eigenvalue calculations]{Estimated CPU time necessary for large matrix eigenvalue calculations}
	\label{cpuAproxEig}
\end{figure}

Similarly, computing the eigenvalues of the companion matrix becomes computationally infeasible as the matrix size grows.  In algorithm analysis, big-O notation, $O(\cdot)$, represents the worst case time complexity of an algorithm as a function of problem size.  For example, the time for an algorithm that is $O(n^2)$ to finish grows quadratically as a function of the problem size, $n$, in the worst case.  Literature has shown that the eigenvalue operation is $O(n^3)$~\cite{linearAlg}.  Therefore, the time to compute the eigenvalue of a matrix can be modeled as a cubic polynomial, with the parameters estimated from measured times.  In the following discussion, all computations were done in MATLAB using Windows 10 64-bit running on an Intel i7-4790K processor at 4.0 GHz with 16 GB of RAM.  The time taken to compute the eigenvalue of a random square matrix was measured as a function of matrix size for a $1\times 1$ matrix to a $7200\times 7200$ matrix.  Then, knowing that the computation time is cubic with respect to matrix size, this result is extrapolated for larger matrix sizes.  Figure~\ref{cpuEig} shows the measured computation times along with the cubic curve fitted to this data.  Figure~\ref{cpuAproxEig} shows the estimated computation times for much larger matrices.

\begin{figure}[htbp]
	\centering
	\includegraphics[width=4in]{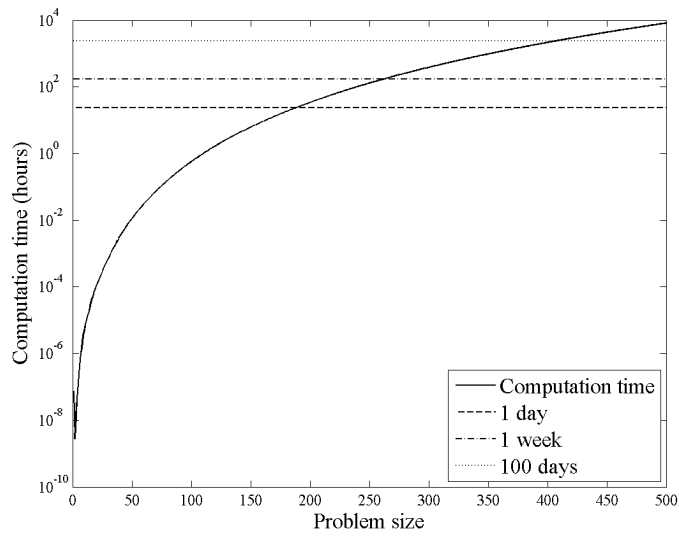}
	\caption[CPU time of Kronecker Multiplication Method]{Computation time needed to compute eigenvalues of companion matrix in Kronecker multiplication method}
	\label{cpuCompanionEig}
\end{figure}

Figure~\ref{cpuCompanionEig} shows the estimated computation time necessary to compute the eigenvalues of the companion matrix in the Kronecker multiplication method based on the curve-fit in Figure~\ref{cpuAproxEig}.

While computation time becomes infeasible for large problem sizes, it is clear that the limiting factor is system memory.  Recall that a problem with 200 states requires 51.2~GB to store the companion matrix in system memory.  From Figure~\ref{cpuCompanionEig}, computing the eigenvalues of said companion matrix is estimated to take one day.  While this is not ideal, it is still more feasible than the required system memory.  

It may be possible to store a large companion matrix as a sparse matrix.  Unfortunately, the sparsity guaranteed by the companion matrix structure is not enough to guarantee that this storage method is more memory efficient than storing the complete matrix.  It is also possible to use iterative numerical methods to compute a subset of the eigenvalues of the companion matrix, but none guarantee that the eigenvalues closest to the imaginary axis are found first.

\subsection{Lyapunov Methods} \label{sub:lyap}
The Rekasius substitution method, Kronecker multiplication method, and other similar root finding methods were primarily developed to analyze small systems ($n<20$).  For larger state-space systems, these methods become computationally infeasible.  Research has been done in analyzing the stability of large time-delayed state space systems, specifically decentralized systems.  These methods typically involve the formulation of a Lyapunov function from system matrices and tuning parameters.  This Lyapunov function is then used to develop an upper-bound for tolerable time delays.  An area of interest in research involves formulating the time delay stability problem as a linear matrix inequality problem.

The main advantage of these methods is that, once derived, it is easy to apply them to any system, and the computations involved are efficient for large systems.  However, the upper-bound produced is typically very conservative compared to the true value of the largest time-delay that the system can tolerate before it first becomes unstable, especially with poor choices for the tunable parameters.  Even with ideal tuning, which can be time consuming, there is no guarantee that the calculated upper-bound is the largest possible.
\section{Methodology}


\subsection{Formulation}

The Rekasius substitution enables a formulation for stability to
be derived that is computationally advantageous.  This derivation
initially introduces the substitution from Eq.~\ref{eq:rekasius}
to the condition for instability from Eq.~\ref{eq:stability}, rather than after symbolically expanding the determinant, to
obtain Eq.~\ref{eq:der1}.  The Rekasius substitution only considers
purely-complex values of $s$ so $1+Ts$ cannot equal 0 and is thus
factored out to
obtain Eq.~\ref{eq:der2} and Eq.~\ref{eq:der3}.  The formulation
in Eq.~\ref{eq:der4} results by noting $T \neq 0$ because the
time delay is non-zero.  That formulation in Eq.~\ref{eq:der4}
is equivalent to the formulation in Eq.~\ref{eq:der5} by noting
the determinant of a block matrix.  Finally, the term of $s$
is separated to derive the expression in Eq.~\ref{eq:der6}.  Note that Eq.~\ref{eq:der6} is an eigenvalue problem for the companion matrix of the polynomial in Eq.~\ref{eq:der3}~\cite{BernsteinMatrix,matrixPoly}.

\begin{eqnarray}
0 & = &  \mathrm{det}\left(sI_n-\left(A_0+A_1\left(\frac{1-Ts}{1+Ts}
\right)\right)\right) \label{eq:der1} \\
& = & \left(\frac{1}{1+Ts}\right)^n\mathrm{det}\left(s(1+Ts)I_n-(1+Ts)A_0-
(1-Ts)A_1\right) \label{eq:der2}\\
& = & \mathrm{det}\left(TI_ns^2+\left(I_n-TA_0+TA_1\right)s-\left(A_0+A_1
\right)\right) = 0 \label{eq:der3}\\
& = & \mathrm{det}\left(I_ns^2+\left(\frac{1}{T}I_n-A_0+A_1\right)s-
\frac{1}{T}\left(A_0+A_1\right)\right) \label{eq:der4}\\
& = & \mathrm{det}\left(\begin{bmatrix} s I_n & -I_n \\ 
-\frac{1}{T}\left(A_0+A_1\right) & s I_n+\left(
\frac{1}{T}I_n-A_0+A_1\right)\end{bmatrix}\right) \label{eq:der5} \\
& = &  \mathrm{det}\left(s I_{2n}-\begin{bmatrix} 0_{n} & I_n \\ 
\frac{1}{T}\left(A_0+A_1\right) & -\left(\frac{1}{T}I_n-A_0+A_1
\right)\end{bmatrix}\right) \label{eq:der6}
\end{eqnarray}

The condition for stability in the presence of a time delay
is thus related to an eigenvalue computation; however, only
the eigenvalues must be purely imaginary to indicate a change
in stability.  Some values of $T$ result in a companion matrix
whose eigenvalues all have a non-zero real part so the system
is never unstable for these values of $T$.  Only values of $T$
resulting in a companion matrix having at least one complex-conjugate
pair of eigenvalues without any real part are associated with
the system becoming unstable.

Finally, similar to the approach in Section~\ref{sub:rek}, the time delay at which the dynamics in Eq.~\ref{eq:dynamics}
become unstable are computed for those values of $T$ at which
some eigenvalues are purely real.  This time delay of $\tau$ results
from the values of $s=\jmath \omega$ and $T$ that satisfy
Eq.~\ref{eq:der6} using the equivalence in Eq.~\ref{eq:tau}.

\subsection{Algorithm}

A search algorithm is utilized to compute the largest value of time
delay for which the dynamics in Eq.~\ref{eq:dynamics} remain stable.
The approach searches over values of $T$ to find instances of
purely-imaginary values of $s$ that satisfy Eq.~\ref{eq:der6} and
then substitute in Eq.~\ref{eq:tau} to find the associated values
of time delay.

The concept of the search depends on the dynamics with $\tau=0$,
and consequently the corresponding system in Eq.~\ref{eq:der1} for $T=0$, being stable.
If the companion matrix for some larger magnitude of $T$ has
poles on the right-hand side, then the companion matrix for
some value between 0 and $T$ must have poles that crossed the
imaginary axis and consequently the dynamics have become unstable
for the associated value of $\tau$.

A pair of searches are actually utilized to find the stabilizing range
of time delay.  The initial search evaluates the companion matrix
at a coarse grid of $T$ to find when the poles of the companion
matrix have changed between the left-hand plane and the right-hand plane.
The final search repeats that evaluation at a fine grid between
any two neighboring values of the coarse grid having the poles changing
planes.  

The set of $T$ values can be limited to a finite set.  This set
should not include $T=0$ because the system is assumed stable
at the value; additionally, the set should not include $T=\infty$
because the companion matrix converges to a constant as $T$
approaches infinity.  The maximum magnitude of $T$ simply
needs to be sufficiently greater than either $\| A_o+A_1\|$
or $\| A_0-A_1\|^{-1}$ so the terms involving $T$ in
the companion matrix are negligibly small compared to the
other terms.

The largest value of time delay for which the dynamics remain
stable is computed by considering those values of $T$ from the set,
along with the associated $\omega$ for each $T$, for which the
companion matrix has some purely-imaginary eigenvalues.
Essentially, a set of $\tau$ values are computed using Eq.~\ref{eq:tau}
using all those $T$ and $\omega$.  The smallest value of $\tau$
from this set is the time delay at which the dynamics in Eq~\ref{eq:dynamics}
become unstable.

\subsection{Computational Characteristics}

\begin{figure}[htbp]
	\centering
	\includegraphics[width=4in]{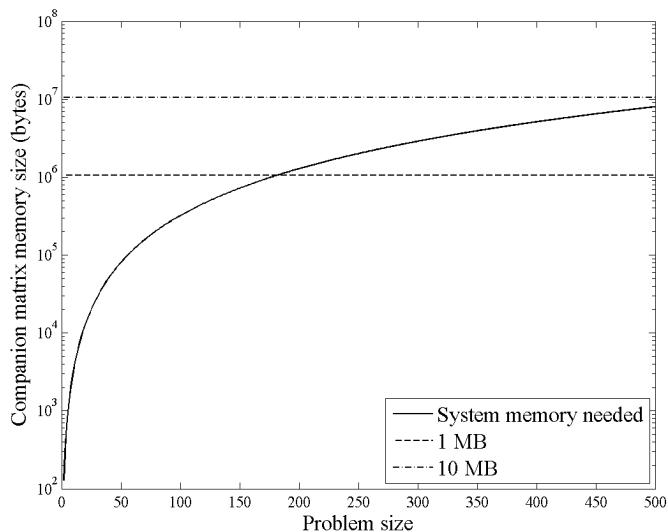}
	\caption[Memory usage of new algorithm]{Memory usage needed to find imaginary axis crossings using the new algorithm as a function of problem size and number of T values searched through}
	\label{noPMem}
\end{figure}

This new algorithm is more memory efficient than the Kronecker multiplication method and other numerical techniques since it solves multiple small problems rather than one large problem.  In each iteration of the algorithm, the companion matrix of this method is only of size $2n$ compared to the $2n^2$ needed by other techniques, given an original state-space of size $n$.  Figure~\ref{noPMem} shows the amount of system memory required to store the companion matrix during each iteration.  Comparing this figure to Figure~\ref{memNeedEig}, it is clear that considerably less memory is required for this algorithm.

The necessary CPU time is also reduced for larger problems, but it is now also dependent on the number of $T$ values used in the search.  Obviously, reducing the search space can reduce the accuracy of the solution and lead to missing root crossing values.  Increasing the search space will increase the computation time linearly.  Furthermore, more accurate results require an even larger search space.  A good strategy may be to perform a coarse resolution search across a large range of $T$ values to find regions of interest (i.e. $|\mathrm{Re}(\lambda)|\leq 10^{-3}$), followed by a finer resolution search within these regions of interest.  

The parallelizability of this approach is also attractive.  The
eigenvalues of the companion matrix at any value of $T$ is computed
independently of any other value of $T$.  The initial overhead
of sharing $A_0$ and $A_1$ data and final overhead of sharing
the set of $\tau$ values is negligible compared to the benefit
of having computation time be a linear function of number of
processors. 

\begin{figure}[htbp]
	\centering
	\includegraphics[width=4in]{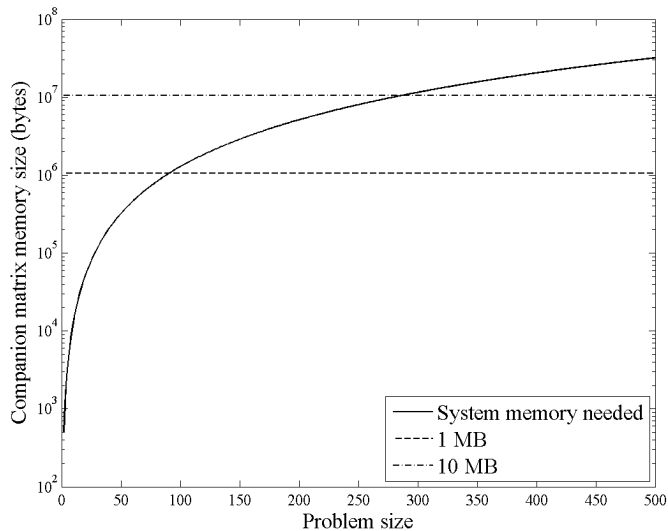}
	\caption[Memory usage of new algorithm with parallel computing]{Memory usage needed to find imaginary axis crossings using the new algorithm when the problem is parallelized across four processors}
	\label{PMem}
\end{figure}

Using parallel processing to reduce the total computation time of the problem comes at the cost of increasing the necessary memory usage.  Specifically, the memory usage increases by a factor of the number of processors used.  In most cases, this increase is negligible compared to the memory usage saved compared to most methods in literature and can be compensated for by dividing the work across multiple machines rather than just multiple processors.  Figure~\ref{PMem} shows the increased memory consumption per iteration if the algorithm is run in parallel on four processors of the same computer.

\section{Examples}
\subsection{Literature example}
The new algorithm is tested with an example from literature to ensure proper operation of the algorithm and assess its accuracy with respect to a problem whose solution is known.  The algorithm is performed on the platform described in Section~\ref{existing}-\ref{sub:kron}.  The problem is characterized by the $A_0$ and $A_1$ matrices shown in Equation~\ref{litExample}~\cite{131}.

\begin{equation}
A_0 = \begin{bmatrix} -1 & 13.5 & -1 \\ -3 & -1 & -2 \\ -2&-1&-4 \end{bmatrix} ; \>\>\> A_1 = \begin{bmatrix}
-5.9 & 7.1 & -70.3 \\ 2 & -1 & 5 \\ 2 & 0 & 6
\end{bmatrix}
\label{litExample}
\end{equation}

The $(T,\omega)$ pairs at which a pole pair crosses the imaginary axis, the corresponding time delays, and the effect on stability are shown in Table~\ref{litResults}.  The smallest time delay at which the system becomes unstable is 0.1624~s.  The system becomes stable again in the range $0.1859<\tau<0.222$.

\begin{table}[htbp]
	\centering
	\caption{Results of Literature Example}\label{litResults}
	\begin{tabular}{cccc}
		\hline \hline
		$T$ (s)   & $\omega$ (rad/s) & Stable/Unstable & Time Delay (s)  \\
		\hline
		-0.4269 & 15.5032 & Unstable & 0.2219, 0.6272, ... \\
		-0.1332 & 0.8407 & Stable & 7.208, 14.682, ...\\
		0.0829 & 3.0347 & Unstable & 0.1624, 2.233, ...\\ 
		0.0953 & 2.9123 & Stable & 0.1859, 2.343, ...\\
		0.6233 & 2.1109 & Unstable & 0.8725, 3.849, ...\\
		\hline \hline
	\end{tabular}
\end{table}

The algorithm reproduces the results shown in literature, with the smallest time delay causing instability as $\tau=0.162~s$.  Based on existing results, an observation is made for applying this algorithm towards a full stability analysis in the presence of time delays.  As $T$ increases near a purely imaginary pole, $s=\jmath\omega$, the direction of the sign change of the real part of $s$ correlates to the effect on stability that the corresponding time delays, $\tau$, have, even though the Rekasius substitution is only valid at $\jmath\omega$.  A pole pair crossing from the left-half plane to the right-half plane correlates to all corresponding time delays having a destabilizing effect, and vice versa. 

\subsection{MAV example}
The second example is a 428-state model of a closed-loop aircraft with a
distributed control and estimation framework.  The details of this model are found in literature and in the addendum~\cite{myGNCpaper}.  The
model contains distributed sensors and actuators along with
distributed processors at nodes throughout the wing.  Some
estimation and control is computed at each node in a decentralized
fashion; however, all sensor data is also transferred through
sequential-bus hardware to the central node in the fuselage for
estimation and control in a centralized fashion.  The bus
introduces an appreciable delay in data transfer to this central
node so the associated stability is of critical importance.  The state dynamics of the delay-free model are given in Eq.~\ref{arrangedStates} (definitions of the submatrices are found in literature), with elements in bold representing sources of time-delay~\cite{myGNCpaper}.

\begin{equation}
\begin{bmatrix} \dot{x}(t)\\\dot{\hat{x}}(t)\\\dot{\hat{x}}^1(t)\\\vdots\\\dot{\hat{x}}^{n_n}(t)\\\dot{x}_{K}^1(t)\\\vdots\\\dot{x}_{K}^{n_n}(t)\end{bmatrix}
=\begin{bmatrix}
A & 0 & 0 & \cdots & 0 & B_1C_{K_1} & \cdots & B_{n_n}C_{K_{n_n}}\\
\boldsymbol{LC} & A-LC & 0 & \cdots & 0 & \boldsymbol{B_{1}C_{K_{1}}} & \boldsymbol{\cdots} & \boldsymbol{B_{n_n}C_{K_{n_n}}}\\
L_{S}^1C_1 & L_{C}^1 & X^1 & \cdots & 0 & BC_K & \cdots & 0\\
\vdots & \vdots &\vdots & \ddots & \vdots & \vdots & \ddots & \vdots\\
L_{S}^{n_n}C_{n_n} & L_{C}^{n_n} & 0 & \cdots & X^{n_n} & 0 & \cdots & BC_K\\
0 & 0 & B_KC & \cdots & 0 & A_K & \cdots & 0\\
\vdots & \vdots &\vdots & \ddots & \vdots & \vdots & \ddots & \vdots\\
0 & 0 & 0 & \cdots & B_KC & 0 & \cdots & A_K
\end{bmatrix}
\begin{bmatrix} x(t)\\\hat{x}(t)\\\hat{x}^1(t)\\\vdots\\\hat{x}^{n_n}(t)\\x_{K}^1(t)\\\vdots\\x_{K}^{n_n}(t)\end{bmatrix}
\label{arrangedStates}
\end{equation}

A trio of existing methods is initially applied to this example. The
number of states is too large for the symbolic computation used
by the existing Rekasius-substitution method since computing the determinant of a $428\times428$ symbolically to obtain the characteristic equation is not feasible.  The
Kronecker-multiplication method is also infeasible for the available
computer because the memory storage exceeds 1~TB of space.  The
only method to compute a valid solution is the Lyapunov method; however,
that method generated an upper bound on stability at a time delay
of $10^{-13}$ seconds that is clearly conservative as shown by
simulations.

\begin{figure}
	\centering
	\includegraphics[width=5in]{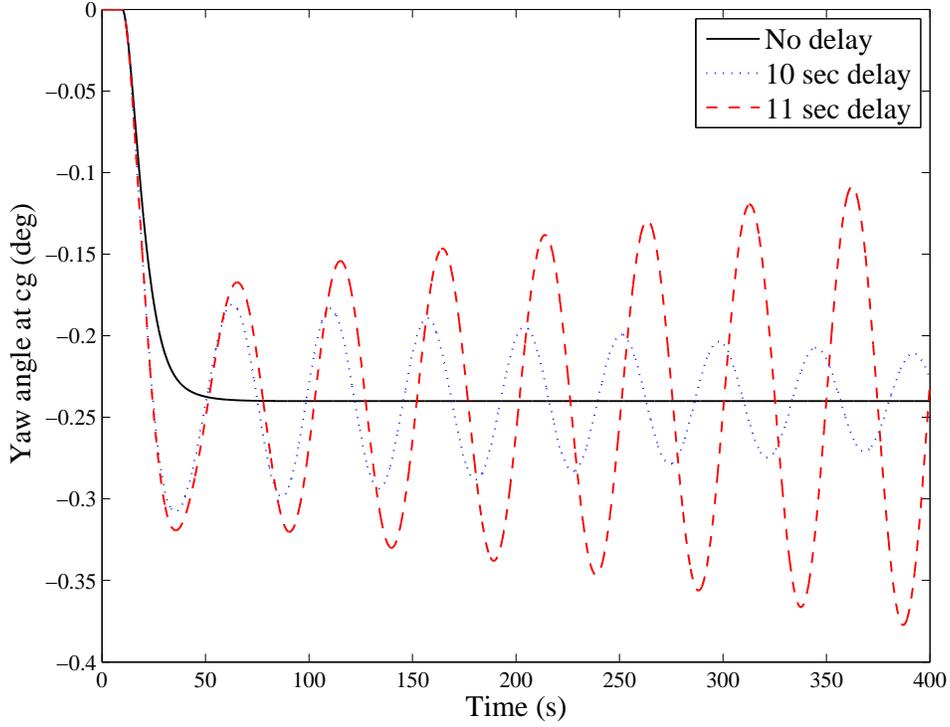}
	\caption[Yaw angle response in presence of time-delay]{Yaw angle of aircraft in response to step command in wing tip twist angle with no time delay, 10~{\em sec} time delay, and 11~{\em sec} time delay}
	\label{fig:yaw}
\end{figure}

The new algorithm is used to analyze the stability of the system by
searching over $T$ values in Equation~\ref{eq:der6} ranging from
-1000~s to 1000~s in increments of 0.001~s and excluding $T=0$.  The
analysis, parallelized across 4 processing cores, is completed in just
over three days of CPU time.  The algorithms computes 10.5~{\em sec}
as the largest value of time delay for which the system remains
stable.  Past this time delay, the lateral-directional dynamics of the system becomes unstable at a frequency of 0.02~Hz.  Simulations show that indeed the system is stable
for a time delay of 10~{\em sec} and unstable for a time delay
of 11~{\em sec}.  To illustrate this instability, Fig.~\ref{fig:yaw} shows the yaw angle at the c.g. of the aircraft in response to a step command to the wing-tip twist angle in the presence of no time delay, 10~{\em sec} time delay, and 11~{\em sec} time delay.

\section{Conclusion}
A new algorithm is presented for the analysis of large state-space systems with time-delays.  The algorithm is adapted from the Rekasius substitution method and uses numerical techniques employed in the Kronecker product method.  Therefore, it maintains the exactness of these techniques while being computationally feasible for large systems.  An analysis is performed that compares the CPU cost and memory consumption of the Kronecker product method with that of the new algorithm.  The parallel nature of the algorithm is used to further reduce computation time necessary for the stability analysis.  The algorithm is used on a system with 428 states, and the results are compared to a Lyapunov-based time-delay algorithm designed for large systems.  It is shown that the largest stable time-delay computed by this new algorithm is much less conservative than the upper-bound obtained by the Lyapunov-based algorithm.

\section{Addendum}

The purpose of this submission is to provide matrix information and code that can be used to recreate the results and compare them to the results in the paper.  The system is an aircraft, $P$, with distributed processors along segments of the wing with local controllers, $K^i$, and local Kalman filters, $F^i$, and a central processor with a central Kalman filter, $F$, receiving delayed inputs.  Subscripts denote the subset of the matrix or vector corresponding to a node, and superscripts denote values computed by that node's processor.

\begin{equation}
P \left\{\begin{matrix}
\dot{x}(t) &=& Ax(t)+\sum_{i=1}^{10}B_iu^i_i(t)\\y(t) &=& Cx(t) 

\end{matrix}\right.
\end{equation}

\begin{equation}
K^i \left\{\begin{matrix}
\dot{x}(t) &=& A_Kx_K(t)+B_K\hat{y}^i(t)\\u^i(t) &=& C_Kx_K(t)

\end{matrix}\right.
\end{equation}

\begin{equation}
F^i \left\{\begin{matrix}
\dot{\hat{x}}^i(t) &=& A\hat{x}^i(t)+Bu^i(t)+L_C^i\left(\hat{x}(t)-\hat{x}^i(t) \right )+L_S^i\left(C_ix(t)-C_i\hat{x}^i(t) \right )
\\y^i(t) &=& C_{14,68}\hat{x}^i(t)

\end{matrix}\right.
\end{equation}

\begin{equation}
F \left\{\begin{matrix}
\dot{\hat{x}}(t) &=& A\hat{x}(t)+\left(\sum_{i=1}^{10}B_iu^i_i(t-\tau) \right )+L\left(Cx(t-\tau)-C\hat{x}(t) \right )

\end{matrix}\right.
\end{equation}

The matrices $A$, $B$, $C$, $A_K$, $B_K$, $C_K$, $L$, $L_S$, and $L_C$ are stored in the file ``matrices.mat".  The script ``delayProof.m" constructs the matrices $A_0$ and $A_1$ corresponding to Equation~\ref{eq:dynamics} for this example and runs the new algorithm described in the paper.  This new algorithm is in the function ``delayStability.m'' and accepts any two square matrices, $A$ and $B$ of the same size as input.  The search grid can be modified in the function.  The output of the function is a two-column matrix whose first column is a list of $T$ values where the companion matrix contains an eigenvalue (pair) near the imaginary axis and the second column contains the corresponding eigenvalue whose imaginary part is positive.  The output of delayProof.m is included for comparison purposes.  The script ``getDelay.m'' performs the second half of the algorithm by processing the matrix output of delayStability.m to perform a finer search approximated as interpolating the imaginary axis crossing points and computing the time-delays associated with each crossing. 

To use the code for a general problem:
\begin{itemize}
	\item Construct $A_0$ and $A_1$ for that problem
	\item Modify the grid defined in delayStability.m as desired
	\item Run the following line: ``crossing = delayStability($A_0$, $A_1$)''
	\item Run getDelay with crossing
\end{itemize}
The delayProof script performs the first three steps for the example shown in the paper.

\end{document}